\documentclass[aps,prb,groupedaddress,twocolumn,superscriptaddress]{revtex4}

\usepackage{cancel}
\usepackage{graphicx}
\usepackage{amssymb, amsmath}
\usepackage{bbold}
\usepackage{float}
\usepackage{multirow}
\usepackage{nicefrac}

\newcommand\bigzero{{\text{\huge0}}}
\newcommand\sz{{\text{\scriptsize 0}}}

\newcommand{\diff}{\mathrm{d}}
\newcommand{\nIm}{\operatorname{Im}}

\newcommand{\cH}{\mathcal{H}}
\newcommand{\cS}{\mathcal{S}}
\newcommand{\cD}{\mathcal{D}}
\newcommand{\cR}{\mathcal{R}}
\newcommand{\tr}[2][\,]{\operatorname{Tr}_{#1}\!\!\left(#2\right)}

\newcommand{\ket}[1]{\vert #1 \rangle}
\newcommand{\ketbra}[2]{\left\vert #1 \middle\rangle\middle\langle #2 \right\vert}
\newcommand{\mean}[1]{\left\langle #1 \right\rangle}

\newcommand{\id}{\mathbb{1}}

\begin{document}

\title{Pitfalls in the theory of carrier dynamics in semiconductor quantum dots:\\ the single-particle basis vs.~the many-particle configuration basis}

\author{T. Lettau}
\affiliation{Institut f{\"u}r Theoretische Physik, Otto-von-Guericke-Universit{\"a}t
  Magdeburg, Postfach 4120, D-39016 Magdeburg, Germany}
\author{H.A.M. Leymann}
\email{leymann@pks.mpg.de}
\affiliation{Institut f{\"u}r Theoretische Physik, Otto-von-Guericke-Universit{\"a}t
  Magdeburg, Postfach 4120, D-39016 Magdeburg, Germany}
\affiliation{Max-Planck-Institut f{\"u}r Physik komplexer Systeme, N{\"o}thnitzer Strasse 38, 01187 Dresden, Germany}
\author{J. Wiersig}
\affiliation{Institut f{\"u}r Theoretische Physik, Otto-von-Guericke-Universit{\"a}t
  Magdeburg, Postfach 4120, D-39016 Magdeburg, Germany}

\date{\today}

\begin{abstract}
We analyze quantum dot models used in current research for misconceptions that arise from the choice of basis states for the carriers. The examined models originate from semiconductor quantum optics, but the illustrated conceptional problems are not limited to this field. We demonstrate how the choice of basis states can imply a factorization scheme that leads to an artificial dependency between two, actually independent, quantities. Furthermore, we consider an open quantum dot-cavity system and show how the dephasing, generated by the dissipator in the von Neumann Lindblad equation, depends on the choice of basis states that are used to construct the collapse operators. We find that the Rabi oscillations of the s-shell exciton are either dephased by the dissipative decay of the p-shell exciton or remain unaffected, depending on the choice of basis states. In a last step we resolve this discrepancy by taking the full system-reservoir interaction Hamiltonian into account.
\end{abstract}

\maketitle

\section{Introduction}
\label{sec:introduction}
There are many well-established theories to describe open quantum many-particle systems consisting of, e.g.,~quasi-free charge carriers in semiconductor heterostructures~\cite{hohenester_density-matrix_1997,hoyer_influence_2003}, cavity photons~\cite{kira_cluster-expansion_2008,mootz_sequential_2012,chow_emission_2014}, phonons~\cite{lorke_influence_2006,kabuss_threshold_2013}, ultracold Bose-gases~\cite{witthaut_beyond_2011,trimborn_decay_2011}, polaritons~\cite{tignon_unified_2000}, and spins~\cite{kapetanakis_spin_2008} in an approximate way. The theories in the preceding references are related to mean-field theories and its successive improvements, like the cluster expansion~\cite{fricke_transport_1996,fricke_improved_1997,hoyer_cluster_2004} (CE), in which equations of motion (EoM) for the mean single-particle occupations and their correlations are derived, whereas higher-order correlations are neglected.

Besides the approximations that are necessary to describe the interacting system itself, many of the referenced models also require additional approximations to include dissipative processes resulting from the systems coupling to an external bath. The von Neumann Lindblad equation is a common procedure to take the influence of the exterior bath on the system into account \cite{breuer_theory_2002,carmichael_dissipation_1999,may_exciton_2003}, provided that the Born-Markov-approximation is justified\cite{nakatani_quantum_2010}.

The experimental progress in the field of cavity quantum electrodynamics in semiconductors \cite{reithmaier_strong_2004,wiersig_direct_2009,nomura_laser_2010,reitzenstein_semiconductor_2012,khitrova_vacuum_2006} shows that there are many interesting systems in which the basic assumption of mean field theories, a large Hilbert space and weak interaction, is not valid. Many of these systems are sufficiently small to be described by their exact wave function or density matrix, formulated in the Hilbert space of all possible many-particle configurations \cite{carmele_antibunching_2010,ritter_emission_2010} without the need for an approximate theory. Despite the efforts that are made to improve the theories from both sides (exact description of relatively small systems \cite{gies_3_2012,carmele_antibunching_2010,richter_numerically_2015}  and approximate description of relatively large systems \cite{leymann_expectation_2014,florian_equation--motion_2013,richter_few-photon_2009,mascarenhas_matrix-product-operator_2015,weimer_variational_2015}), there is still a gap between those systems that are small enough to be described exactly and those that are large enough to fulfill the requirements of approximate theories like the CE.

In this article, we describe pitfalls in the choice of the basis states that may occur, when applying approximate theories on small systems within or close to the mentioned gap. We use three examples to contrast approaches that are based on a formulation in single-particle states with approaches that use many-particle configuration states as a basis.
Although the formulations are equivalent, the choice of basis states can decide about further steps. In our three examples, we show that the choice of basis states can suggest misleading approximations or determine the modeling of dissipative processes, which leads to deviations of the results, in the two formulations that go beyond simple approximation errors.

The remainder of this paper is organized in the following way: In Sec.~\ref{sec:issue}, we discuss, based on an extended Jaynes-Cummings model \cite{jaynes_comparison_1963,shore_jaynes-cummings_1993} (JCM), the effects of a fallacious mean-field factorization scheme, implied by a description in single-particle states.

In Sec.~\ref{sec:lt}, we consider an open system treated in the von Neumann Lindblad (vNL) formalism. We demonstrate that the basis states in which the collapse operators and with it the dissipator of the vNL equation are constructed can actually influence the modeling of the system. In the first example concerning the vNL (Sec.~\ref{ssec:hole}), the choice of basis states limits the possibilities to adjust the model to the experimental situation. Whereas, in the second example concerning the vNL formalism (Sec.~\ref{ssec:ssd}), the basis states determine whether two parts of the system are affected by the environment independently or intertwined and one system part is dephased by the dissipative decay of the other. Finally we recapitulate how the dissipator in the vNL can be constructed from a system plus reservoir approach\cite{carmichael_dissipation_1999} and resolve the misconception, that has led to results depending on the choice of basis states (Sec.~\ref{ssec.sytempres}).

The last section \ref{sec:last} summarizes and concludes the paper. In the appendix, we give details of the EoM and the parameter space of the semiconductor JCM (App.~\ref{app:sjcm}). Furthermore we outline the derivation of the analytic solution for the open system (App.~\ref{app:deph}), and present the effects of an additional external pump on the open system (App.~\ref{app:pump}).

\section{Fallacious factorization}
\label{sec:issue}
To illustrate a conceptual problem that can arise from a Hartree-Fock-like factorization of expectation values, we consider a model with Jaynes-Cummings interaction, introduced in \cite{richter_few-photon_2009}, with the Hamiltonian
\begin{align}
	H =& \omega b^\dagger b + \varepsilon_e e^\dagger e + \varepsilon_h h^\dagger h - (g h e b^\dagger +  \mathrm{h.c.}),
\end{align}
where $b^{(\dagger)}$ annihilates (creates) a cavity mode photon with frequency $\omega$ and $e^{(\dagger)}/h^{(\dagger)}$ annihilates (creates) an electron/hole, with energy $\varepsilon_e/\varepsilon_h$, respectively. The dipole matrix element $g$ can be chosen real, and all parameters are specified in units of $\hbar$.
This Hamiltonian describes a two-level quantum dot (QD) embedded in a semiconductor environment, coupled to a single cavity mode. It would be identical to the JCM, if one restricts the electronic states to fully correlated electrons and holes, i.e.,~restricting the electronic states to a single exciton (electron-hole pair). However, in order to describe the semiconductor properties of a QD, an independent occupation of the electron and hole states is allowed in this model, which we term the semiconductor JCM. Both systems perform a coherent exchange between the cavity photons and the exciton, called Rabi oscillations \cite{shore_jaynes-cummings_1993}. The calculation of the time evolution using EoM for the expectation values produces a hierarchy of coupled equations. We will show that a factorization of many-particle expectation values into single-particle expectation values, often used to truncate hierarchies of EoM, is not only unnecessary in this exemplary model, but also leads to conceptually wrong conclusions.

To derive the EoM, we follow the approach of \cite{richter_few-photon_2009}, in which the photons are not described by creation and annihilation operators, but by the photon probability distribution. This allows for a variant of the CE in which the photonic part is treated exactly, and is termed the photon probability CE by the authors of \cite{richter_few-photon_2009}. The expectation values of interest are the hole $f_h=\mean{h^\dagger h}$ and the electron $f_e=\mean{e^\dagger e}$ occupation, the occupation of the Fock-states with $n$ photons $p_n = \mean{\ketbra{n}{n}}$, and the imaginary part of the photon-assisted polarization \mbox{$\psi_n = \nIm\mean{\ketbra{n+1}{n}h e}$}. From the Heisenberg equation for the generalized occupations $f^{e}_n = \mean{\ketbra{n}{n}{e^\dagger e}}$ and $f^{h}_n = \mean{\ketbra{n}{n}{h^\dagger h}}$, with $f^{e/h} = \sum_n f^{e/h}_n$, we obtain the time derivatives
\begin{align}
	\diff_t f^{e/h}_n =& 2g\sqrt{n+1}~\psi_n\label{eq:time_fn},\\
	\diff_t p_n = &2g\sqrt{n+1}~\psi_n	- 2g \sqrt{n}~\psi_{n-1},\label{eq:time_pn}\\
	\diff_t\psi_n =	  &-g\sqrt{n+1}~(p_{n+1}-f^h_{n+1} -f^e_{n+1})\nonumber\\
												   -& g\sqrt{n+1}~\left(C_{n+1}^X - C_n^X \right),	\label{eq:time_psin}
\end{align}
where the diagonal terms are zero since the cavity is chosen to be in resonance with the QD. The EoM for the photon-assisted polarization couples to the higher-order term
\begin{align}
	C_n^X =  \big\langle \ketbra{n}{n} e^\dagger e h^\dagger h \big\rangle
\end{align}
that describes the electron-hole correlation. In the JCM, in which electrons and holes are perfectly correlated, the many-particle term $C_n^X$ can be expressed exactly by the already known single electron expectation values $f_n^e$, thus closing the hierarchy. In a semiconductor environment, the assumption of perfectly correlated electrons and holes is not valid~\cite{berstermann_correlation_2007}. Therefore, the term $C_n^X$ can no longer be expressed by a single single-particle term.

\paragraph*{Guided by the single-particle basis,} one can proceed with an approximate treatment of $C^X_n$. The first order of the photon probability CE results in the factorization
\begin{align}
  \begin{split}
      C_n^X = \mean{\ketbra{n}{n}e^\dagger e h^\dagger h } &\approx \frac{\mean{\ketbra{n}{n} e^\dagger e}\mean{\ketbra{n}{n}h^\dagger h}}{\mean{\ketbra{n}{n}}}  \\
      &= \frac{f^e_nf_n^h}{p_n},
  \end{split}
\end{align}
which is related to a neglect of the electron-hole correlation
\begin{align}
      \delta=\mean{e^\dagger e h^\dagger h} - \mean{e^\dagger e}\mean{h^\dagger h}\label{eq:purecorr},
\end{align}
and corresponds to the Hartree-Fock approximation. With the applied factorization one obtains a closed set of EoM and is able to calculate the dynamics of the electronic and photonic occupations $N = \mean{b^\dagger b} = \sum np_n$, and the photon autocorrelation function~\cite{glauber_quantum_1963} at zero delay time $\tau= 0$
\begin{align}
	g^{(2)}(0) = \frac{\mean{b^\dagger b^\dagger b b}}{\mean{b^\dagger b}^2} = \frac{\sum (n^2 - n)p_n}{(\sum n p_n)^2}.
\end{align}

\begin{figure}[]
	\centering
    \includegraphics[width=0.48\textwidth]{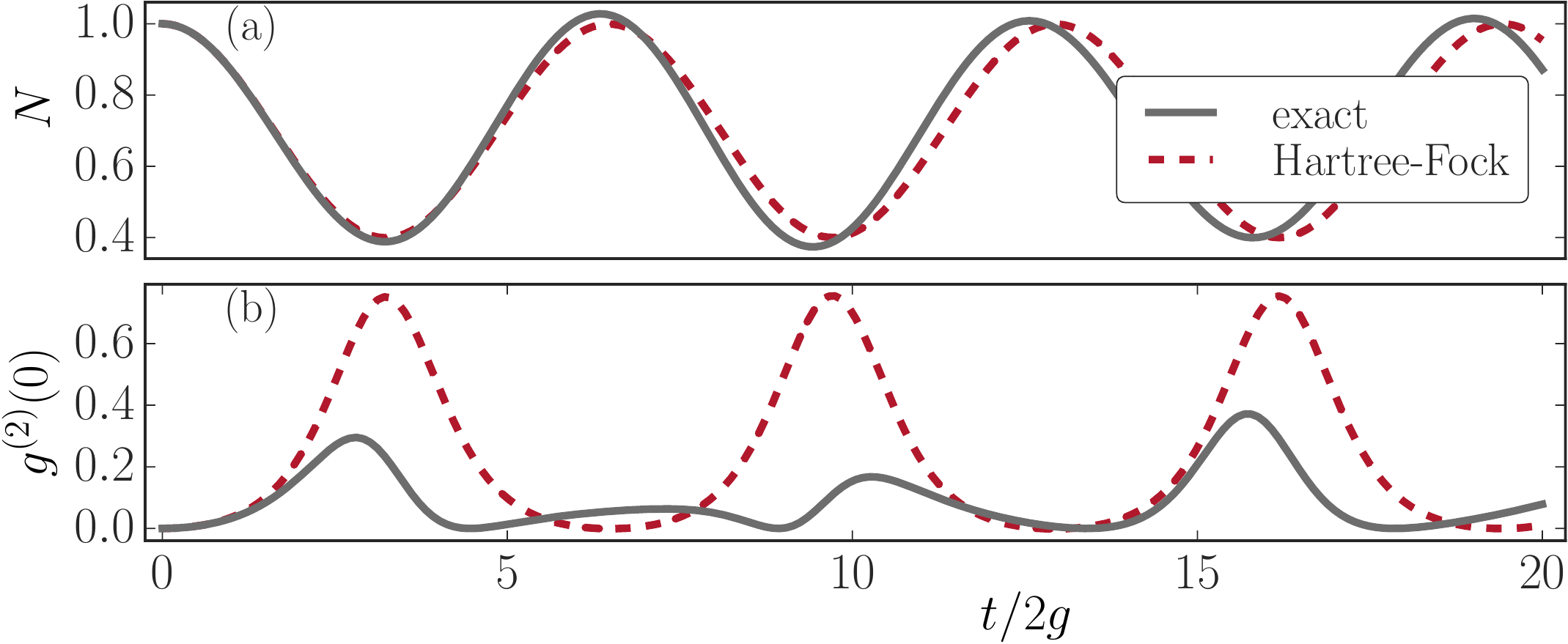}
  \caption{Time evolution of the semiconductor JCM with exact and factorized EoM (Hartree-Fock), for the mean photon number $N$ (a) and the photon autocorrelation $g^{(2)}(0)$ (b); Initial conditions: $p_1=1$, $f^e, f^h = 0.3, 0.1$.}
  \label{fig:eom}
\end{figure}

\begin{figure}[]
\centering
    \includegraphics[width=0.48\textwidth]{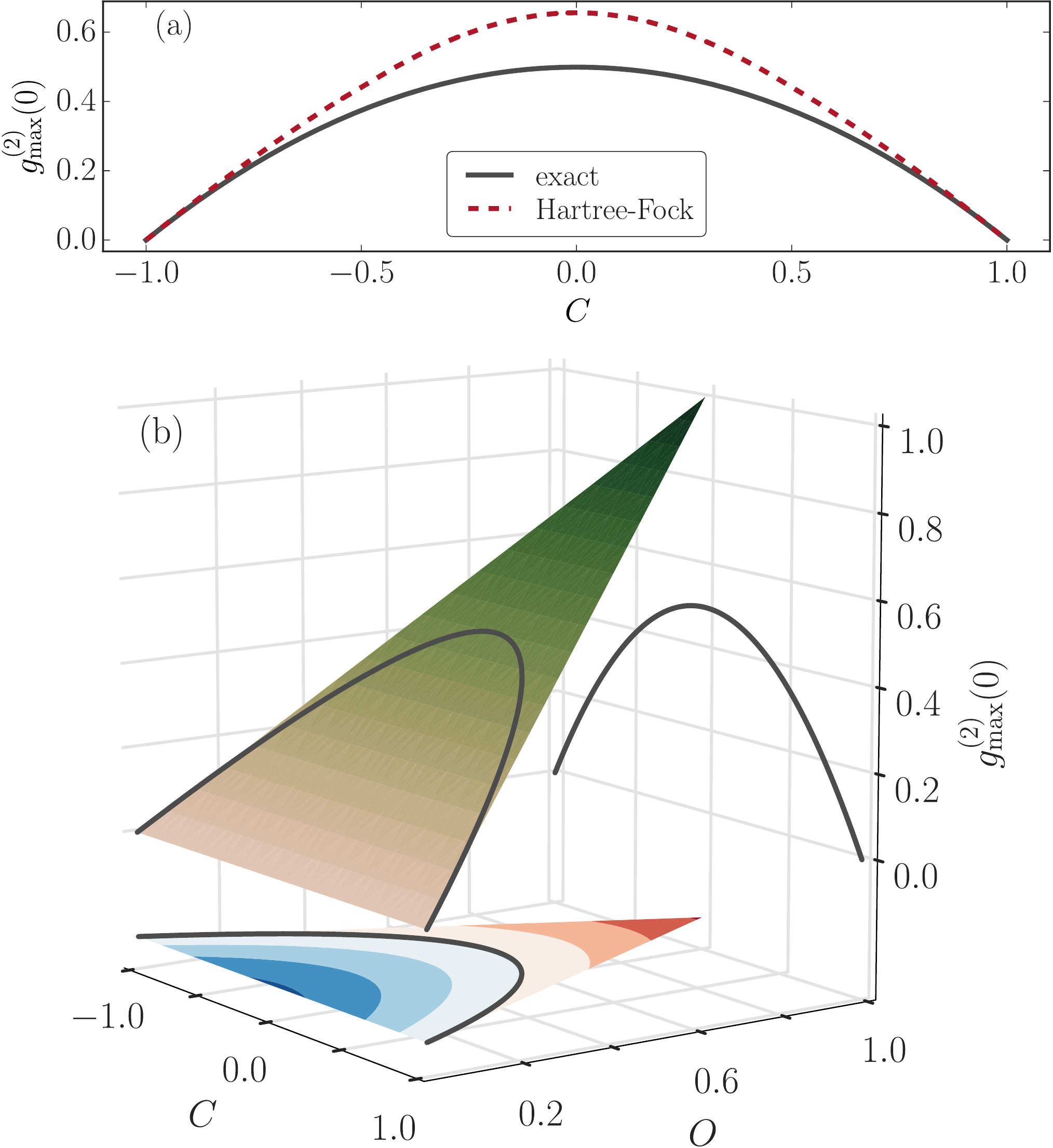}
  \caption{The dependence of the maximum amplitude of $g^{(2)}(0)$ on the charge $C$ and the oscillation ability $O$. (a): $g^{(2)}_\mathrm{max}(0)$ in dependence of $C$ for the exact and factorized (Hartree-Fock) EoM. (b): $g^{(2)}_\mathrm{max}(0)$ (green area) in dependence of $C$ and $O$, and the electron-hole correlation $\delta$ (blue/red contour plot in the $C$-$O$-plane) which increases with $O$ from $\delta=-\nicefrac{1}{4}$ to $\delta=\nicefrac{1}{4}$. The special case of $\delta=0$ is marked by the black curves. The initial conditions are $\delta=0$, $f^e+f^h=1$ and $p_1=1$. To fix the additional free parameter we have chosen the initial probabilities for $\ket{G}$ and $\ket{X_s}$ to be equal, which is equivalent to the restriction to a fixed number of total excitations.}
  \label{fig:correlation}
\end{figure}

Figure~\ref{fig:eom} shows the dynamics of the semiconductor JCM for a cavity with an initially prepared single-photon Fock state and $f^e,f^h = 0.3,0.1$. The dimensionless charge $C = f^h - f^e$ of the QD is preserved by the Hamiltonian. It is proposed in \cite{richter_few-photon_2009} that, with a fixed number of total initial excitations (i.e.,~$p_n=const$ and $f^e+f^h=const$), the charge $C$ determines the maximum amplitude of the photon autocorrelation function $g^{(2)}_\mathrm{max}(0)$. Figure \ref{fig:correlation}(a) shows the dependence of $g^{(2)}_\mathrm{max}(0)$ on $C$, when the system is initially prepared in a single-photon Fock state $p_1=1$ and $f^e+f^h=1$ for the exact (see next paragraph) and the factorized system. The curves have their maximum at $C=0$, which suits the notion that the probability that an electron can recombine with a matching hole, i.e,~the ability of the system to oscillate, is directly connected to $C$. The exact and the approximate curve deviate, since the electron-hole correlation $\delta$ is forced to be zero for all times in the factorized version of the EoM, whereas $\delta=0$ is only an initial condition in the exact EoM (see next paragraph).

\paragraph*{When the system is described in many-particle configurations,} it becomes apparent that the previous conclusion is only an artifact of the focus on single-particle properties, which manifests in the factorization of the term $C_n^X$\footnote{Actually calculating the time derivative of $C_n^X$ reveals that it couples only to the known quantities $\psi_n$ (App.~\ref{app:sjcm}).}. Factorizing the term $C_n^X$ forces the electron-hole correlation $\delta$ to be zero and introduces a constraint to the system that eventually results in the artificial connection between $g^{(2)}_\mathrm{max}(0)$ and $C$.
\begin{figure}[]
\centering
  \includegraphics[width=0.4\textwidth]{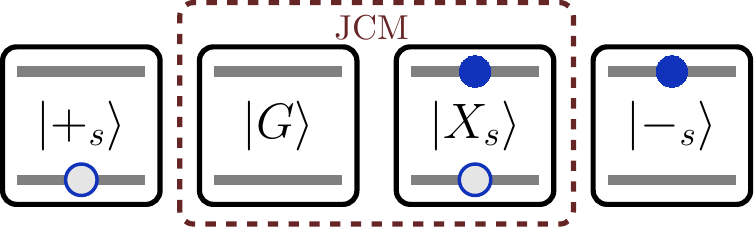}
  \caption{Illustration of the electronic configurations $\ket{i}$ of the semiconductor JCM. The original JCM consists of the states within the dashed box, which are the only ones appearing in the interaction part of the Hamiltonian in Eq.~(\ref{eq:sjcm_config}).}
  \label{fig:configs}
\end{figure}
A reformulation in terms of many-particle configurations, depicted in Fig.~\ref{fig:configs}, reveals which electronic states of the QD take part in the Rabi oscillations. The state of the system is determined by four coefficients $c_i=\mean{\ketbra{i}{i}}$ ~\footnote{In the incoherent regime all considered observables, only depend on the absolute value of the coefficients. So w.l.o.g.\ they can be chosen real. The free choice of the coefficients is reduced by condition $\operatorname{Tr}\rho=1$, resulting in three independent coefficients. In the formulation using creation and annihilation operators the three degrees of freedom are $f^e$, $f^h$ and $C^X$.}, corresponding to the many-particle configuration states $\ket{i}$. Examining the Hamiltonian formulated in this basis
\begin{align}
        H &= \omega b^\dagger b + \sum \varepsilon_i  \ketbra{i}{i} -\left(g\ketbra{G}{X_s}b^\dagger + \mathrm{h.c.}\right)\label{eq:sjcm_config}
\end{align}
reveals that only $\ket{G}$ and $\ket{X_s}$ take part in the Rabi oscillations. This finding suggests the definition of a new quantity, the oscillation ability $O = \mean{\ketbra{G}{G}} + \mean{\ketbra{X_s}{X_s}}$, which actually determines the amplitude of the Rabi oscillations, rather than the charge $C = \mean{\ketbra{+_s}{+_s}} - \mean{\ketbra{-_s}{-_s}}$. Even in the case of $C=0$ and fixed excitations, one could have no Rabi oscillations at all, if the initial electronic state of the system is equally distributed between the configurations $\ket{+_s}$ and $\ket{-_s}$. Figure~\ref{fig:correlation}(b) shows $g^{(2)}_\mathrm{max}(0)$ in dependence of $O$ and $C$ for a constant amount of total excitations (see App.~\ref{app:parameter}). The amplitude of $g^{(2)}(0)$ increases with $O$, \emph{independent} of $C$\,~\footnote{The domain of $O$ is determined by $C$ (Eq.~(\ref{eq:domain}), Fig.~\ref{fig:correlation}~(b)) and therefore, the maximum amplitude of $g^{(2)}(0)$ is indirectly affected by $C$.}. The correlation between electrons and holes $\delta$ is depicted as a contour plot at the bottom of Fig.~\ref{fig:correlation} (b), varying from anticorrelated to fully correlated electrons and holes with increasing $O$. The special case for the Hartree-Fock factorization $\delta=0$ is marked by the three black curves. Following this path in parameter space, one regains the artificial dependence of the maximum amplitude of $g^{(2)}(0)$ on the charge $C$. This dependence, projected in the $C$-$g^{(2)}(0)$-plane, is identical to the black curve in Fig~\ref{fig:correlation}~(a).
\paragraph*{In conclusion,} we have demonstrated that in the semiconductor JCM, not the charge of the QD $C$ but the oscillation ability $O$ determines the maximum of $g^{(2)}$. The constraint in configuration space introduced by the factorization scheme can lead to a misconception about the systems dynamic, in our case it is the connection between the charge of the QD and its ability to perform Rabi oscillations. The effect of this constraint in configuration space is especially drastic in our case, since a connection between observables of the system is derived, $g^{(2)}_{max}(0)=f(O,C)$, in contrast to a case where the dependence of an observable on an external parameter is derived, e.g.\ the input-output characteristics $\mean{b^{\dagger}b}=f(Pump_{\textnormal{ext}})$ of a laser.

One can avoid problems and misconceptions like this by describing the finite states of the carriers localized in the QD in the basis of its many-particle configurations as we have demonstrated here. When a formulation in single-particle states is desirable one should include all correlations between the localized single-particle states [App.~\ref{app:sjcm}] since correlations between single-particle states are strong in finite systems \cite{leymann_expectation_2014}. There are many approaches on QD-(cavity) systems described in the literature, that either find a formulation that includes all possible many-particle configurations of the system \cite{ritter_emission_2010,richter_numerically_2015} or if this is not possible use hybrid factorization schemes related to the cluster expansion. These factorization schemes are hybrid approaches in the sense that the correlations between the carriers localized in a QD are fully included, while correlations between other system parts are treated approximately by factorization e.g.\ correlations between, different QDs \cite{leymann_sub-_2015,jahnke_giant_2016,foerster_computer-aided_2017}, QD and delocalized wetting-layer states \cite{kuhn_hybrid_2015}, and QD states and continuum states of the light-field in free space \cite{florian_equation--motion_2013}.

\section{Open systems and the construction of the dissipator}
\label{sec:lt}
In the previous section, we have discussed misleading results that arise from an approximation scheme that truncates the hierarchy of EoM.
In this section we demonstrate that when open quantum system are described in Born-Markov approximation using the
vNL equation
\begin{align}
	\diff_t \rho = &i[\rho,H] + \sum_i \gamma_i \left(L_i \rho L_i^\dagger -\frac{1}{2} L_i^\dagger L_i \rho -\frac{1}{2} \rho L_i^\dagger L_i\right) \nonumber\\
			  =& i[\rho,H] + \cD (\rho)\label{eq:lindblad}
\end{align}
it can make a significant difference whether the dissipator $\cD$ is constructed in a single-particle or in a many-particle configuration basis.
We demonstrate that a misleading assumption can already be incorporated in the construction of the EoM, thus producing questionable results even if the basic EoM for $\rho$ is then solved without further approximations.

\subsection{Hole capture}
\label{ssec:hole}

As a first introductory example, we consider the hole capture of a semiconductor QD. To model the hole capture from delocalized wetting layer states, the model illustrated in Fig.~\ref{fig:configs} is augmented by further localized states. For cylindrical QDs these states are the p-shell states, which are energetically higher than the s-shell states. Restricting the model to one spin direction and one state in the p-shell results in four single-particle states that can be occupied by up to four carriers. This model, consisting of 16 possible many-particle configurations (see Ref.~\cite{gies_3_2012} for details), is the basis for many models used to describe semiconductor QDs \cite{ritter_emission_2010,leymann_sub-_2015,jahnke_giant_2016}.

\begin{figure}[]
\centering
    \includegraphics[width=0.47\textwidth]{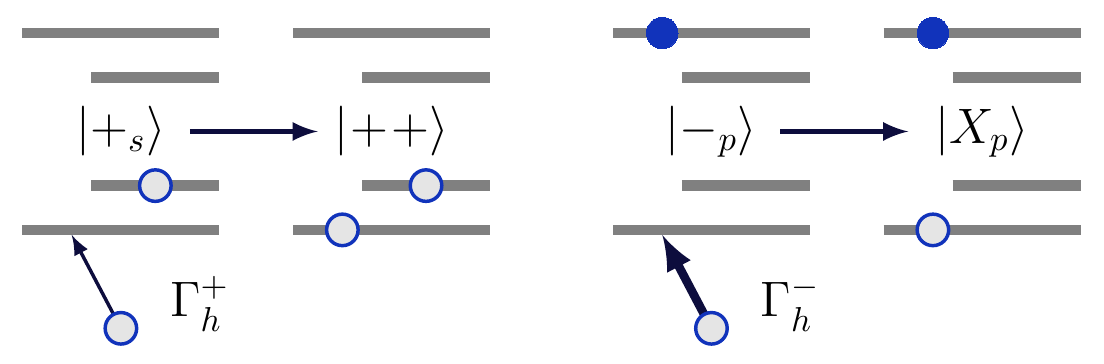}
    \caption{Illustration of the transition from the many-particle configuration $\ket{+_s}$ to \mbox{$\ket{\!+\!+}$} and $\ket{-_p}$ to $\ket{X_p}$, corresponding to the collapse operators $L_1$ and $L_2$, respectively. Both transitions result in the capture of a hole in p-shell of the QD. The capture rate of a positively charged hole depends on the QD's charge ($\Gamma_h^+ < \Gamma_h^-$).}
  \label{fig:capture}
\end{figure}

The excitation of the QDs is facilitated by electron and hole capture from the quasi-continuous wetting layer states into the p-shell.
To describe the hole capture in the single-particle basis, one uses a single collapse operator, $L = h_p^\dagger$ in Eq.~\eqref{eq:lindblad}, that creates a hole in the p-shell. Assigned to this process is a hole capture rate $\Gamma_h$. This formulation treats the hole capture in the p-shell independently of the occupation of the other states. 
However, the carriers are captured due to phonon and Coulomb scattering of the delocalized wetting layer carriers into the localized QD states and the single-particle-energies of the QD states are renormalized by the Coulomb interaction. Since the scattering rates depend on the energies of the final state, the hole capture rate of a positively charged QD is lower than the one of a negatively charged QD \cite{steinhoff_treatment_2012}, as illustrated in Fig.~\ref{fig:capture}.
 To model the hole capture in a way that takes different capture rates into account, one needs to construct a collapse operator for each transition between two many-particle configurations in which a hole is created, with rates depending on the configurations. Two exemplary transitions are illustrated in Fig.~\ref{fig:capture}, which create a hole in the p-shell, and correspond to the operators $L_1 = \ketbra{++}{+_s}$ and $L_2 = \ketbra{X_p}{-_p}$, with the rates $\Gamma_h^+ < \Gamma_h^-$, respectively.

This example illustrates two possible ways to construct a transition of a carrier, triggered by the environment, within the dissipator $\cD$: (i) using single-particle creation and annihilation operators, resulting in a single collapse operator (e.g.,~$L = h_p^\dagger$ for the hole capture). (ii) using a set of different collapse operators formulated as transition operators between many-particle configurations (e.g.,~$L_1$ and $L_2$). This formulation allows for a direct distinction between different many-particle configurations.

Note that the dissipator in (i) can be also obtained using configuration operators, $L=h_p^\dagger = \sum_{ij}\ketbra{i}{j}$ (with $i,j$ chosen so that $L$ creates a p-shell hole). Accordingly, a combination of creation and annihilation operators can regain a distinction between the configurations as in (ii). However, these alternative ways would result in a rather clumsy notation.
\paragraph*{The conclusion} of this introductory example is that  the dissipator $\cD$ can be constructed in two diffrent ways and that it appears necessary to formulate the dissipator  in the basis of many-particle configurations.
\subsection{Non-local dephasing}
\label{ssec:ssd}
In this example, we show that the two ways to construct the dissipator $\cD$, described in Sec.~\ref{ssec:hole}, lead to different results even when the rates for the different collapse operators formulated in the many-particle basis, are equal. We emphasize that the same set of operators is used in both constructions of $\cD$
and that the only difference is how the operators enter the dissipator.

\begin{figure}[]
\centering
    \includegraphics[width=0.485\textwidth]{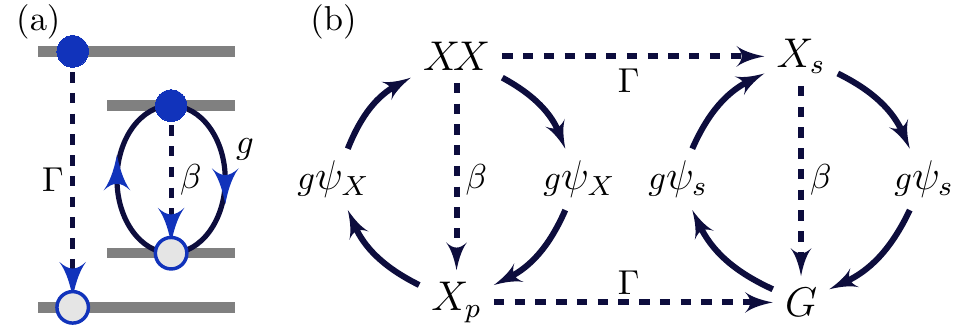}
    \caption{(a): Illustration of the system dynamics in the single-particle basis. (b): Illustration of the transitions between the many-particle configurations revealing that there are two Rabi cycles, connected by the decay of the p-exciton.}
  \label{fig:decay}
\end{figure}

\begin{figure}[]
\centering
    \includegraphics[width=0.47\textwidth]{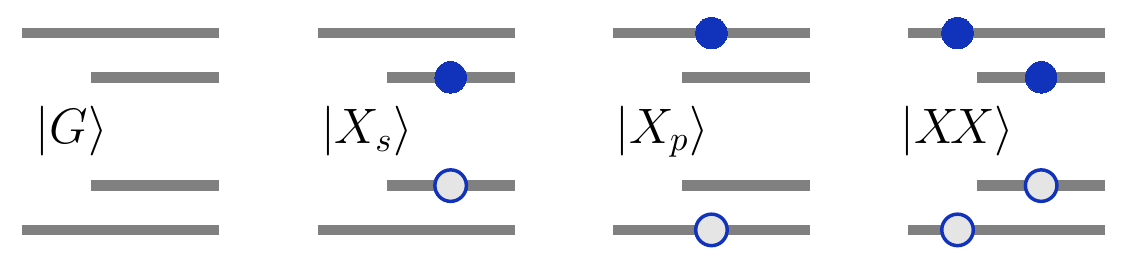}
    \caption{Illustration of the many-particle configurations $\ket{G}$, \mbox{$\ket{X_s}$}, $\ket{X_p}$, and $\ket{X\!X}$, which are the basis states for the QD-model exhibiting non-local dephasing.}
  \label{fig:all_states}
\end{figure}

Such a situation is the vacuum Rabi oscillation of an electron-hole pair in the s-shell in resonance with a high quality cavity mode, in presence of the spontaneous decay of an electron-hole pair in the p-shell, as illustrated in Fig.~\ref{fig:decay} (a). The basis states of the Hilbert space for this system are $\ket{n,i}$, where $n$ is the number of cavity photons and $i$ denotes the electronic configuration of the QD. With an initially empty cavity and a QD prepared in the biexciton state, four electronic configurations are coupled by the vNL equation: The ground state configuration $\ket{G}$, the s-exciton $\ket{X_s}$, the p-exciton $\ket{X_p}$ and the biexciton $\ket{X\!X}$ configuration, as illustrated in Fig.~\ref{fig:all_states}.
The Jaynes-Cummings interaction Hamiltonian reads
\begin{align}
    H_{JC} = &-\left(gh_s e_sb^\dagger + \mathrm{h.c.}\right) \nonumber\\
           = &-\left(g\left(\ketbra{G}{X_s} + \ketbra{X_p}{X\!X}\right)b^\dagger + \mathrm{h.c.}\right)
\end{align}
in the single-particle and the configuration basis respectively.
The dissipator generates the spontaneous loss of excitons in the p- and s-shell, with the rates $\Gamma$ and $\beta$, respectively. In contrast to the hole capture in Sec.~\ref{ssec:hole}, the decay rates in the p-shell are independent of the oscillatory state of the s-exciton. Unlike the Hamilton operator, which is independent of the formulation, the effect of the dissipator $\cD$ depends on its formulation since the collapse operators enter nonlinearly. In the single-particle basis, the loss of the p-shell exciton is generated by $L_{\textrm{sp}} = h_p e_p$ (formulation (i)). The same operator can be constructed by a sum of configuration operators
\begin{align}
L_{\textrm{sp}} = \ketbra{G}{X_p} + \ketbra{X_s}{X\!X},
\label{eq:Lsp}
\end{align}
which is still formulation (i). In the many-particle formulation, the spontaneous loss of p-shell excitons is generated by two collapse operators
\begin{align}
L_{G} = \ketbra{G}{X_p}\quad \textnormal{and}\quad L_{X} = \ketbra{X_s}{X\!X},
\label{eq:LGLX}
\end{align}
with equal rates $\gamma_{G} = \gamma_{X} = \Gamma$ (formulation (ii)). The same holds for the spontaneous exciton loss in the s-shell, with the loss rate $\beta$ and the collapse operators chosen accordingly.
\paragraph*{In the single-particle basis,} the dynamics of the s- and the p-shell are decoupled, which can be seen in the EoM for the single-particle operator expectation values
\begin{align}
\begin{split}
	\diff_t \mean{e^\dagger_p e_p} &= -\Gamma	\mean{e^\dagger_p e_p},\\
	\diff_t \mean{e^\dagger_s e_s} &= -\beta	\mean{e^\dagger_s e_s} +\underbrace{2g\psi}_{\mathrm{Rabi}^+},\\
		\diff_t \psi &= -\beta\psi +\underbrace{g \left(\mean{b^\dagger b } - \mean{e_s^\dagger e_s}\right)}_{\mathrm{Rabi}^-},\\
	\diff_t \mean{b^\dagger b }&=\hspace{12ex}-\underbrace{2g\psi}_{\mathrm{Rabi}^+},
\end{split}\label{eq:rabi}
\end{align}
with $\psi$ being the imaginary part of photon-assisted polarization ($\psi = \nIm{\mean{h_se_sb^\dagger}}$) and $\mathrm{Rabi}^{\pm}$ marking the terms responsible for the Rabi-oscillations. The p-shell occupation decays exponentially with rate $\Gamma$, the s-shell occupation oscillates with the vacuum Rabi-frequency $2g$ and decays with rate $\beta$, and the polarization is subject to the dephasing introduced by the spontaneous losses $\beta$ in the s-shell. Fig.~\ref{fig:decay} (a) illustrates the dynamics of the single-particle occupations of the system.

\paragraph*{In the configuration basis,} the required quantities to formulate the EoM are the occupations of the basis states ($X\!X^n$, $X^n_p$, $X^n_s$, $G^n$) with photon number $n$, e.g.,~$G^n=\mean{\ketbra{n,G}{n,G}}$ and the photon-assisted polarizations between bi- and p-exciton $\psi_X^n = \nIm(\mean{\ketbra{n,X\!X}{n+1,X_p}})$ and between s-exciton and ground state $\psi_s^n = \nIm(\mean{\ketbra{n,X_s}{n+1,G}})$. Since we start with an empty cavity, the EoM are restricted to the first photon block $(n=0,1)$ and read
\begin{align}
\begin{split}
	\diff_t X\!X^0&= -(\Gamma +\beta)X\!X^0 +2g\psi^0_X ,\\
	\diff_t \psi^0_X&=     -g X\!X^0  -(\Gamma+\nicefrac{\beta}{2})\psi^0_X	+ g X^1_p,\\
	\diff_t X^1_p&= -2g \psi^0_X	   -\Gamma X^1_p ,\\
	\diff_t X^0_s&=  \Gamma X\!X^0		 -\beta X^0_s		+2g \psi^0_s,\\
	\diff_t \psi^0_s&= \{\Gamma,0\}\psi^0_X	  -g X^0_s  - \nicefrac{\beta}{2}\psi^0_s	+ g G^1,\\
	\diff_t G^1&= \Gamma  X^1_p  -2g\psi^0_s,\\
	\diff_t X^0_p&= \beta	X\!X^0	 -\Gamma X^0_p,\\
	\diff_t G^0&= \beta X^0_s	    	  +\Gamma X^0_p.\\
\end{split}\label{eq:diffconfig}
\end{align}
The curled brackets $\{\Gamma,0\}$ in the fifth line of Eqs.~\eqref{eq:diffconfig} mark the difference between the single-particle \mbox{(i:~$\Gamma$)} and the configuration basis (ii:~$0$) in the EoM, which we will discuss in more detail below. For further discussion it is convenient to formulate the EoM in matrix form $\diff_t r = Mr$, where the column vector $r=\left( X\!X^0,\dots,G^0\right)^T$ contains the dynamical quantities as listed in Eqs.~\eqref{eq:diffconfig} (for initial state $r_0=(1,0,\dots,0)^T$) and the parameter matrix $M$ reads
\begin{align}
   M= \left(\begin{array}{ccc|ccc|cc}
          -\Gamma -\beta & 2g					&    \sz  			&      					\\
        -g & -\Gamma-\nicefrac{\beta}{2}	& g &   &    &  \multicolumn{1}{c}{\bigzero}   \\
            \sz			 &  -2g					&   -\Gamma   		&     					\\\hline
            \Gamma		 &  \sz						&   \sz   			&      -\beta		& 2g			&   \sz  			&  	 \sz  &   \sz \\
           \sz			 &  \{\Gamma,0\}				&   \sz   			&  -g & - \nicefrac{\beta}{2}	& g	&  	 \sz  &   \sz \\
            \sz			 &  \sz						&  \Gamma  			&   \sz				&  -2g			&   \sz 			&  	 \sz  &   \sz \\\hline
            \beta		 &  \sz 					&   \sz   			&   \sz			    &  \sz				&   \sz   			& -\Gamma &   \sz \\
            \sz			 &  \sz						&   \sz   			&  \beta	    	&  \sz 				&   \sz  			&  \Gamma &   \sz \\
    \end{array}\right). \label{eq:diffmatrix}
\end{align}
The matrix $M$ can be separated into eight blocks, indicated by the lines in Eq.~(\ref{eq:diffmatrix}). We refer to these blocks row-by-row. Block I describes the Rabi oscillations with frequency $2g$ on its off-diagonal elements and the decay of excitation and the dephasing of the polarization on its diagonal elements. The same holds for block IV and the off-diagonal elements of these blocks correspond to the terms $\mathrm{Rabi^\pm}$ in Eq.~(\ref{eq:rabi}). Since there is no pumping in this system, block II, which transports occupation from lower to higher electronic states, is zero. Block III together with the diagonal elements of block I describes the transfer of population from the part of the system with a p-shell exciton to the one without. The occupation that is lost due to the negative sign of the $\Gamma$s in block I is transferred to occupations without a p-shell exciton by the positive $\Gamma$s of block III. The entry in curled brackets $M_{\psi_s,\psi_X}=\{\Gamma, 0\}$ reflects the difference between the construction of the dissipator in single-particle basis (i) ($M_{\psi_s,\psi_X}=\Gamma$) and in configuration basis (ii) ($M_{\psi_s,\psi_X}=0$). In the  single-particle basis, the photon-assisted polarization is transferred from the $X\!X$ - $X_p$ oscillation to the $X_s$ - $G$ oscillation. Therefore, only the s-exciton decay with rate $\beta$ and not the p-exciton decay with rate $\Gamma$ contributes to the dephasing of $\psi$ in Eq.~(\ref{eq:rabi}). The loss of polarization $\psi^0_X$ with rate $\Gamma$ is transferred to $\psi^0_s$ with exactly the same rate. On the contrary in the many-particle basis, the polarization $\psi^0_X$, which is lost by the decay of the p-shell exciton, is not picked up by the polarization $\psi^0_s$, thus the element $M_{\psi_s,\psi_X}$ is zero. The remaining blocks can be interpreted analogously, by associating pairs of positive and negative entries in the same column with a transfer of occupation. The transitions between the states are illustrated in Fig.~\ref{fig:decay} (b).

\begin{figure}
\centering
    \includegraphics[width=0.48\textwidth]{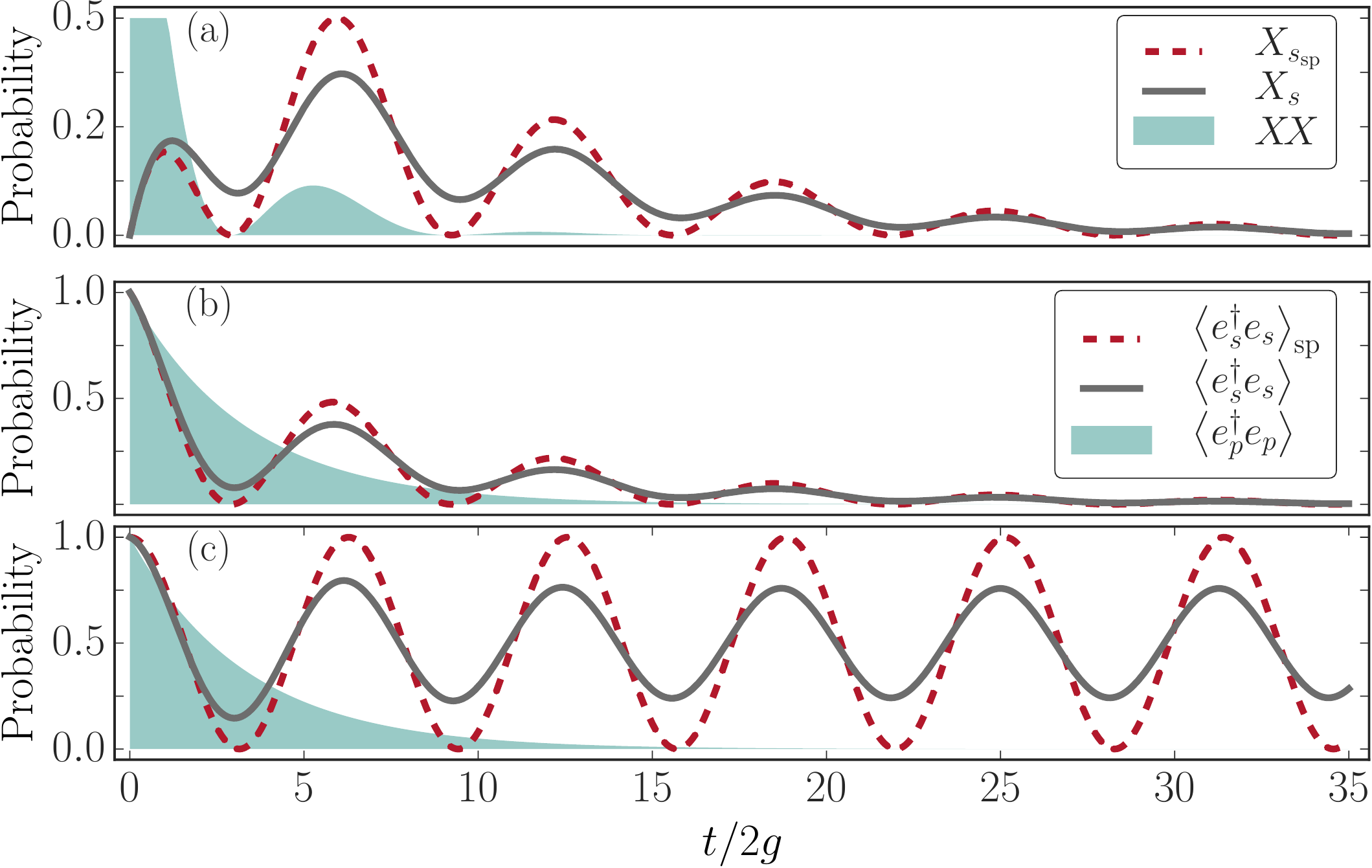}
    \caption{Dynamics of the s-shell Rabi oscillations and spontaneous p- and s-shell decay obtained with a the dissipator constructed in the single-particle basis (i) ('sp', red dashed line) and in the many-particle basis (ii) (black line). (a): Occupation probability of the s-exciton $X_s$, the shaded area marks the biexciton occupation $X\!X$. (b, c): Occupation probability of s-shell electron occupation $\mean{e^\dagger_se_s}$, the shaded area marks the p-shell electron occupation $\mean{e^\dagger_pe_p}$. The decay rates are $\beta=0.25$ and $\Gamma = 0.3$ measured in units of the Rabi frequency $2g$ for panel (a) and (b), in panel (c) the rate $\beta=0$.}
  \label{fig:dephasing}
\end{figure}

\paragraph*{The solutions of the numerical integration} of Eq.~(\ref{eq:diffconfig}) are shown in Fig.~\ref{fig:dephasing}. In panel (a) and (b), a generic case for the time evolution of the system, for the configuration probabilities in (a) and the single-particle occupations in (b), is depicted. Here the deviations of the two approaches are visible but one might overlook or dismiss them as irrelevant.
The results for the occupation of the s-exciton state $\ket{X_s}$ and s-shell electron $\mean{e^\dagger_s e_s}$ depend on the construction of the dissipator. The results obtained in the single-particle basis are labeled by the subscript 'sp'.
The initially prepared biexciton (panel (a), shaded area) oscillates with the Rabi frequency $2g$ and decays with the rate $\Gamma +\beta$. The s-exciton occupation increases with rate $\Gamma$, oscillates with the Rabi frequency, and decays with the rate $\beta$. This behavior holds for both, the construction of the dissipator in the many-particle configurations and in the single-particle basis. The two curves deviate in the fact that in the single-particle basis, the oscillations have a larger amplitude than in the many-particle basis, and that in the single-particle basis, the ground state is fully occupied within each Rabi cycle. An alternative representation of the dynamics is given in panel (b), in which the single-particle expectation values for the electrons $\langle e^\dagger_ie_i\rangle$ are shown. The p-electron is decaying with rate $\Gamma$ in both formulations of the dissipator, whereas the oscillation of the s-shell electron depends on the formulation of the dissipator.
To emphasize the characteristic difference between the two constructions of the dissipator we consider the limiting case of vanishing s-shell decay ($\beta = 0$), where the deviations are not blurred by a circumstantial dephasing mechanism, with $\langle e^\dagger_ie_i\rangle$ shown in panel (c). In the single-particle basis, the s-shell performs Rabi oscillations with a constant amplitude of \nicefrac{1}{2}, while the p-shell exciton decays exponentially. In the many-particle basis, the p-shell electron decay dephases the Rabi oscillations in the s-shell, resulting in a diminished amplitude in the long-term behavior. Due to the construction of the dissipator in the non-local basis of the many-particle configurations the dissipation in one system part induces non-local dephasing in an otherwise independent system part.

\begin{figure}[]
    \includegraphics[width=0.48\textwidth]{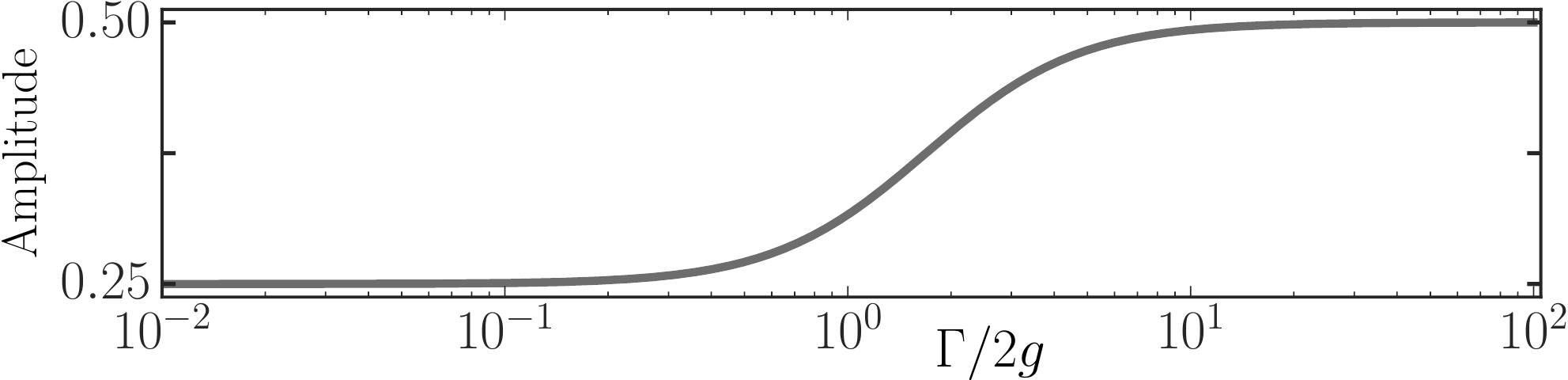}
    \caption{Asymptotic effect of the non-local dephasing on the amplitude of the oscillation of $\mean{X_s}$ in dependence of the scaled decay rate $\tilde{\Gamma}$.}
  \label{fig:amplitude}
\end{figure}

In many cases, e.g.~in cw-lasers, the long term behavior or the steady state of the system are of interest. The simple form of the EoM in the case of vanishing $\beta$ allows to derive analytic expressions for the dependence of the amplitude of the Rabi oscillations on the rate $\Gamma$ (see App.~\ref{app:deph} for details). In the long term behavior the amplitude of the Rabi oscillations $A$ can be expressed by
\begin{align}
	A|_{t\gg \frac{1}{\Gamma}} = \frac{1}{2}\frac{\sqrt{(\tilde{\Gamma}^2+2)^2 + \tilde{\Gamma}^2}}{(\tilde{\Gamma}^2+4)},\quad \tilde{\Gamma} = \frac{\Gamma}{2g}.
\end{align}
Note the peculiar result that the long term effect of the non-local dephasing is strongest, when its rate is the smallest since in this case the Rabi oscillations are exposed to the dephasing for the longest time. As it can be seen in Fig.~\ref{fig:amplitude}, the minimal amplitude is $\nicefrac{1}{4}$ for almost vanishing but nonzero decay rates $\tilde{\Gamma}$. In the opposite case of an immediate p-exciton decay, the amplitude remains at its maximum value of $\nicefrac{1}{2}$ since no polarization could build up to be dephased.

Going beyond this minimal example one can further increase the non-local dephasing effect by exploiting the same mechanism discussed above. Adding an additional pump process to the dissipator $\cD$, compensating the p-shell loss, ties the s-exciton permanently to the dephasing influence of the p-shell. In this case the Rabi oscillations in the s-shell would completely vanish, when the dissipator is constructed in the configuration basis (ii). Whereas when the dissipator is constructed of in the single-particle basis (i), the Rabi oscillations in the s-shell would again not be effected at all by the p-shell (see App.~\ref{app:pump}).

The problematic conclusion to this section is that the outcome of the EoM depends crucially on the choice of basis states for constructing the dissipator. In the next section we will see how this problem can be resolved and that, in contrast to our first example in Sec.~\ref{sec:issue}, the non-local dephasing effect is not an artifact of an approximation error.
\subsection{System plus reservoir approach}
\label{ssec.sytempres}
The discrepancies between the results, when the dissipator $\cD$ is constructed in either single-particle (i) or the configuration basis (ii) originate from deviating approximations and assumptions about the system-reservoir interaction, already build into the construction of the dissipator $\cD$ itself. To see where the crucial assumptions deviate we discuss in this section how the dissipator describing the decay of a p-shell exciton in Sec.~\ref{ssec:ssd} can be derived from a system plus reservoir approach.

Starting from the von Neumann equation $\diff_t \chi=i[\chi,H]$ for the full density operator $\chi$ describing the QD-cavity-mode system and a reservoir of non-confined modes, we derive the EoM for the reduced density operator $\rho=\tr[\cR]{\chi}$ in Born-Markov approximation \cite{carmichael_dissipation_1999}. To this end we divide the Hilbert space $\cH$ into a reservoir part $\cH_{\cR}$ consisting of the non-confined modes and a system part $\cH_{S}=\cH_{QD}\otimes\cH_{C}$ consisting of the QD and the confined cavity mode. The QD Hilbert space itself consists of the s- and p-shell subspace $\cH_{QD}=\cH_{s}\otimes\cH_{p}$. After recapitulating how one can derive the general EoM for $\rho$, where we essentially follow the approach from Ref.\cite{carmichael_dissipation_1999}, we compare the obtained EoM (formulated in the single-particle and in the configuration basis) to the EoM for $\rho$ used in the previous section.

Assuming a reservoir of harmonic modes with frequency $\omega_k$ that are annihilated(created) by $ r_k^{(\dagger)}$,  we can formulate the reservoir Hamiltonian $H_{\cR}$ and the system-reservoir interaction Hamiltonian $H_{\cS\Leftrightarrow \cR}$ as
\begin{align}
    H_{\cR} =&\sum_{k}\omega_k r_k^{\dagger}r_k,\\
    H_{\cS\Leftrightarrow \cR}=& \sum_{j}\left(\sum_{k}\kappa^{j}_{k} r_k^{\dagger}L_{j}+\sum_{k}\kappa^{j*}_{k} r_k L^{\dagger}_{j}\right)\nonumber\\
       =&\sum_j\left(R^{\dagger}_j L_j+R_j L^{\dagger}_j\right)
   \label{eq:systemreshamilt}
\end{align}
respectively. In $H_{\cS\Leftrightarrow \cR}$ the sum over all reservoir modes is summarized in the reservoir operators $R_j$ coupling to the system operators $L_j$ in full rotating wave approximation\cite{carmichael_dissipation_1999,nakatani_quantum_2010}. The operators $L_j$, will be chosen as $L_{\textrm{sp}}$ according to Eq.~\eqref{eq:Lsp} in the single-particle (i) and as $L_{G}$ and $L_{X}$ according to Eq.~\eqref{eq:LGLX} in the configuration basis formulation (ii).

In Born approximation the full density operator $\chi(t)$ factorizes to $\chi(t)=\rho(t)\otimes \rho^T_\cR$, where $\rho^T_\cR$ is the reservoir density operator in thermal equilibrium. We trace over the reservoir $\cR$ and reformulate the von Neumann equation in the interaction picture for $\chi(t)=\rho(t)\otimes \rho^T_\cR$ as an integro-differential equation
\begin{align*}
&\diff_t\rho(t)=\int_0^t dt^\prime\tr[\cR]{[H_{\cS\Leftrightarrow \cR}(t),[\rho(t^\prime) \rho^T_\cR,H_{\cS\Leftrightarrow \cR}(t^\prime)]]},
\end{align*}
describing the dissipative influence of the reservoir $\cR$ on the reduced density operator $\rho$. Now we insert the general Hamiltonian from Eq.~\eqref{eq:systemreshamilt} and execute the commutators and collect all reservoir operators in the reservoir correlations $\tr[\cR]{\bullet\,\rho^T_\cR}=\mean{\bullet}_\cR$. When the reservoir occupations can be neglected the only contributing reservoir correlations are $\langle R_j(t^\prime)R^{\dagger}_i(t)\rangle _\cR$ and the EoM for the reduced density operator reads
\begin{align*}
&\diff_t\rho(t)=\sum_{i,j}\int_0^tdt^\prime\mean{R_j(t^\prime)R^{\dagger}_i(t)}_\cR\Big\lbrace L_i(t)\rho(t^\prime) L^{\dagger}_j(t^\prime)\\
&-L^{\dagger}_j(t^\prime)L_i(t)\rho(t^\prime)+L_i(t)\rho(t^\prime) L^{\dagger}_j(t^\prime) -\rho(t^\prime) L^{\dagger}_j(t^\prime)L_i(t)\Big\rbrace .
\end{align*}
When the time scales of the reservoir and the system can be separated we can apply the Markov approximation, which corresponds to
\begin{align}
\label{eq:rescorrelations}
\mean{R_j(t^\prime)R^{\dagger}_i(t)}_\cR=&\sum_{kl}\delta_{kl}\kappa^{i\ast}_l\kappa^{j}_ke^{i\omega_k t^\prime}e^{-i\omega_l t}\\
=&\sum_k\kappa^{i\ast}_k\kappa^{j}_ke^{-i\omega_k (t-t^\prime)}\approx\gamma_{ji}\delta(t-t^\prime),\nonumber
\end{align}
and we obtain
\begin{align}
&\diff_t\rho=\widetilde{\cD}(\rho)=\sum_{i,j}\gamma_{ji}\Big\lbrace2L_i\rho L^{\dagger}_j-L^{\dagger}_jL_i\rho-\rho L^{\dagger}_jL_i\Big\rbrace.
\label{eq:dissipatorsystempbath}
\end{align}
Here the dissipator $\widetilde{\cD}$ has a more general non-diagonal form \cite{breuer_theory_2002} in the collapse operators $L_i$ and rates $\gamma_{ij}$, in contrast to the dissipator $\cD$ in Eq.~\eqref{eq:lindblad} used in Sec.~\ref{ssec:ssd}.
This non-diagonal dissipator appears in many systems, e.g.,~in open resonators the non-diagonal form of the dissipator induces correlations between different photon modes\cite{hackenbroich_field_2002,hackenbroich_quantum_2003,eremeev_quantum_2011,fanaei_effect_2016}.
We now use the non-diagonal dissipator $\widetilde{\cD}$ from Eq.~\eqref{eq:dissipatorsystempbath} and insert the system operators $L_j$ from the system-reservoir interaction Hamiltonian formulated in the single-particle basis and in the configuration basis.
\paragraph*{In the single-particle basis} the system-reservoir interaction Hamiltonian reads
\begin{align*}
 H^{\mathrm{sp}}_{\cS\Leftrightarrow \cR}=&\sum_{k}\kappa_{k} r_k^{\dagger}h_p e_p+\sum_{k}\kappa_{k} r_k  e^{\dagger}_ph^{\dagger}_p\\
       =&R^{\dagger}_{sp} L_{sp}+R_{sp} L^{\dagger}_{sp}\\
       =&\sum_{j=sp}\left(R^{\dagger}_j L_j+R_j L^{\dagger}_j\right)
\end{align*}
with $\kappa_{k}$ being the coupling strength of reservoir mode $k$ to the p-exciton. This Hamiltonian leads to the dissipator
\begin{align}
\widetilde{\cD}_{\mathrm{sp}}(\rho)=&\Gamma\Big\lbrace2L_{sp}\rho L^{\dagger}_{sp}-L^{\dagger}_{sp}L_{sp}\rho-\rho L^{\dagger}_{sp}L_{sp}\Big\rbrace,
\label{eq:spsystempbath}
\end{align}
where we have identified the only appearing rate $\gamma_{\textrm{sp}\textrm{sp}}$ with the rate $\Gamma$ from the previous section. Equation~\eqref{eq:spsystempbath} is identical to the dissipative part of the EoM~\eqref{eq:lindblad} used in Sec.~\ref{ssec:ssd} in single-particle formulation (i) with $L_j=L_{\mathrm{sp}}$.
Using Eq.~\eqref{eq:Lsp} for $L_{sp}$ and $L^{(\dag)}_{G}L^{(\dag)}_{X}=0$ we can reformulate Eq.~\eqref{eq:spsystempbath} to
\begin{align}
\begin{split}
\widetilde{\cD}_{\mathrm{sp}}(\rho)&=\Gamma\Big\lbrace2L_{G}\rho L^{\dagger}_{G}-L^{\dagger}_{G}L_{G}\rho-\rho L^{\dagger}_{G}L_{G}\Big\rbrace\\
&+\Gamma\Big\lbrace2L_{X}\rho L^{\dagger}_{X}-L^{\dagger}_{X}L_{X}\rho-\rho L^{\dagger}_{X}L_{X}\Big\rbrace\\
&+2\Gamma L_{G}\rho L^{\dagger}_{X}+2\Gamma L_{X}\rho L^{\dagger}_{G},
\end{split}\label{eq:spsystempbathinconf}
\end{align}
which corresponds to the $\Gamma$ dependent part of Eqs.~\eqref{eq:diffconfig} and \eqref{eq:diffmatrix} with $M_{\psi_s,\psi_X}=\Gamma$.
\paragraph*{In the configuration basis} the system-reservoir interaction Hamiltonian reads
\begin{align*}
\begin{split}
    H^{\mathrm{C}}_{\cS\Leftrightarrow \cR}=&\sum_{k}\kappa^{G}_{k} r_k^{\dagger}L_{G}+\sum_{k}\kappa^{X}_{k} r_k^{\dagger}L_{X}\\
    &+\sum_{k}\kappa^{G*}_{k} r_k  L_{G}^{\dagger}+\sum_{k}\kappa^{X*}_{k} r_k L_{X}^{\dagger}\\
       =&R^{\dagger}_{G} L_{G}+R^{\dagger}_{X} L_{X}+R_{G} L^{\dagger}_{G}+R_{X} L^{\dagger}_{X}\\
       =&\sum_{j=G,X}\left(R^{\dagger}_j L_j+R_j L^{\dagger}_j\right)
       \end{split}
\end{align*}
where we have allowed the dipole-matrix elements $\kappa^{j}_{k}$ to depend on s-exciton state $j=G,X$, which would not be possible in the single-particle basis. By inserting the system operators operators $L_j$ into Eq.~\eqref{eq:dissipatorsystempbath} we obtain
\begin{align*}
\begin{split}
\widetilde{\cD}_{\mathrm{C}}(\rho)&=\gamma^{C}_{GG}\Big\lbrace2L_{G}\rho L^{\dagger}_{G}-L^{\dagger}_{G}L_{G}\rho-\rho L^{\dagger}_{G}L_{G}\Big\rbrace\\
&+\gamma^{C}_{XX}\Big\lbrace2L_{X}\rho L^{\dagger}_{X}-L^{\dagger}_{X}L_{X}\rho-\rho L^{\dagger}_{X}L_{X}\Big\rbrace\\
&+2\gamma^{C}_{XG}L_{G}\rho L^{\dagger}_{X}+2\gamma^{C}_{GX}L_{X}\rho L^{\dagger}_{G}.
\end{split}
\end{align*}
This dissipator $\widetilde{\cD}_{\mathrm{C}}$ is in general not in agreement with the diagonal dissipator $\cD$ from Eq.~\eqref{eq:lindblad}.
For rates $\gamma^{C}_{ij}=\Gamma$ the dissipator $\widetilde{\cD}_{\mathrm{C}}$ agrees with the dissipator constructed in the single-particle basis $\widetilde{\cD}_{\mathrm{sp}}$ in Eq.~\eqref{eq:spsystempbathinconf}.
If we assume the system-reservoir coupling strength to be independent of the s-shell exciton $\kappa^{G}_{k}=\kappa^{X}_{k}=\kappa_{k}$, we obtain $\gamma^{C}_{ij}=\Gamma$ and thus $\widetilde{\cD}_{\mathrm{C}}=\widetilde{\cD}_{\mathrm{sp}}$. In fact in this case the system-reservoir interaction Hamiltonians are identical with $H^{\mathrm{C}}_{\cS\Leftrightarrow \cR}=\id_s\otimes H^{\mathrm{sp}}_{\cS\Leftrightarrow \cR}$, where $\id_s$ is the identity operator in $ \cH_s$. This resolves the problematic conclusion from Sec.~\ref{ssec:ssd} and we see that starting from the system-reservoir interaction Hamiltonian leads to a dissipator that is in general non-diagonal and independent from the choice of basis states \footnote{An analog approach, based on a system-reservoir interaction Hamiltonian, is also advisable to describe the hole capture from our introductory example in Sec.~\ref{ssec:hole}.}.

\begin{figure}[]
\centering
    \includegraphics[width=0.48\textwidth]{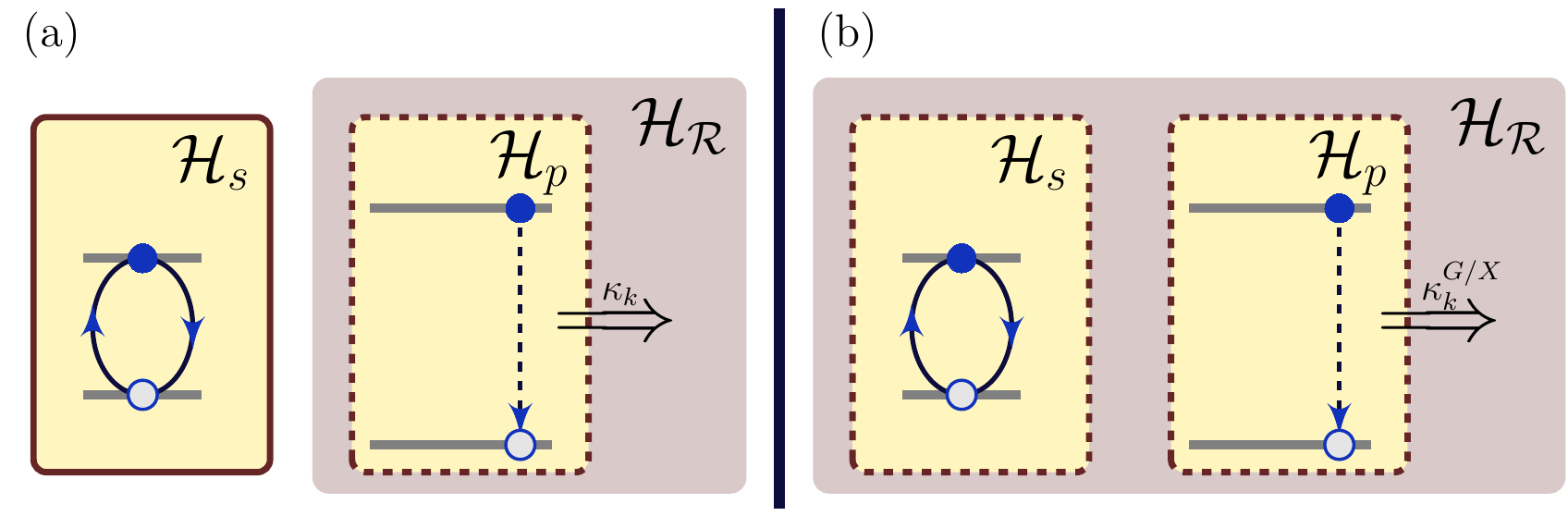}
    \caption{Illustration of the different reservoir couplings. In the left figure (a) the reservoir coupling elements $\kappa$ are independent of the state of s-exciton. The interaction Hamiltonian $H^{\mathrm{sp}}_{\cS\Leftrightarrow \cR}$ operates in $\cH_p\otimes\cH_\cR$, thus the reservoir interacts only with a single localized state. In the right figure (b) the reservoir coupling elements $\kappa^j$ depend on the state of s-exciton thus $H^{\mathrm{C}}_{\cS\Leftrightarrow \cR}$ operates in $\cH_s\otimes\cH_p\otimes\cH_\cR$ and the p-exciton loss is connected to a non-local measurement of the s- and p-exciton state corresponding to $L_{G/X}$.}
  \label{fig:localvsnonlocalres}
\end{figure}

 When we use the diagonal form of the dissipator ad-hoc as done in Eq~\eqref{eq:lindblad}, we implicitly make strong assumptions about the reservoir, namely that the reservoir correlations result in rates
\begin{align}
 \gamma^{C}_{GG}=\gamma^{C}_{XX}=\Gamma \quad \textnormal{and}\quad\gamma^{C}_{XG}=\gamma^{C}_{GX}=0.
 \label{eq:sprratesforspooky}
\end{align}
Nevertheless, from a formal point of view it is possible to construct a reservoir Hamiltonian that leads to the rates in Eq.~\eqref{eq:sprratesforspooky} and thus the described non-local dephasing effect. To this end it is however necessary that the coupling strengths $\kappa_k^j$ depend on the s-exciton state and thus the system-reservoir Hamiltonian interacts non-locally with the QD \cite{schirmer_stabilizing_2010,may_exciton_2003} as illustrated in Fig.~\ref{fig:localvsnonlocalres}.
\section{Conclusion}
\label{sec:last}
We have shown how the choice of basis states can change the dynamics of a system, if an approximation is involved in the calculation. In our first example, the appearance
of the equations, formulated in a single-particle basis, suggested a factorization scheme, which created an artificial dependence between two actually independent quantities.
We have analyzed this dependence in terms of the systems many-particle basis states, in which the relations between the quantities can be seen directly.

In the second part, we have investigated an open system treated in Born-Markov-approximation, where the reservoir influence is modeled by a dissipator in Lindblad form.
We have shown that the way, in which an equal set of collapse operators enter the dissipator, has a profound influence on the systems dynamics. The construction of the dissipator determines
if the Rabi oscillations of the s-shell exciton are non-locally dephased by the decay of the p-shell exciton.

The problem of formulation dependent dynamics, has been resolved, by taking the system-reservoir interaction Hamiltonian into account.
Starting from the full Hamiltonian and evaluating the reservoir correlation functions, we have shown that in both formulations, the s-shell Rabi
oscillations are independent of the p-shell decay. However we also shown that the non-locally dephased s-shell oscillations can actually occur when the system-reservoir interaction Hamiltonian depends on the whole QD state. In contrast to the first
example, the misconception in second part arises not from an inappropriate approximation scheme, but from the notion that two differently constructed dissipators would
describe the same physical situation.

\section{Acknowledgments}

We thank T.~Pistorius for a stimulating discussion, that has led to our last example. We would also like to thank the two unknown referees who provided us with very helpful hints for improvements and constructive criticism. T. Lettau and H.A.M. Leymann have contributed equally to this work.

\newpage
\appendix

\section{Semiconductor JCM}
\label{app:sjcm}
\subsection{Equation of Motion}
The EoM for the photon-assisted polarization can be obtained by
\begin{widetext}
    \begin{align}
    \begin{split}
	\diff_t \mean{h e \ketbra{n+1}{n}]}	 =& \mean{i[H, g h e \ketbra{n+1}{n}]} \\
					=&-ig \mean{ \underbrace{\cancel{h e b^\dagger h e \ketbra{n+1}{n}}}_{\propto ~h h} + e^\dagger h^\dagger b h e \ketbra{n+1}{n}  - \underbrace{h e \ketbra{n+1}{n}\cancel{h e b^\dagger}}_{\propto ~h h} + \ketbra{n+1}{n} e^\dagger h^\dagger b} \\
				 =&  -ig\sqrt{n+1}\mean{e^\dagger e h^\dagger h\ketbra{n}{n} - ee^\dagger hh^\dagger \ketbra{n+1}{n+1} }\\
				 =&  -ig\sqrt{n+1}\langle e^\dagger e h^\dagger h\ketbra{n}{n} - (1-e^\dagger e)(1-h^\dagger h)\ketbra{n+1}{n+1} \rangle\\
				 =&  ig\sqrt{n+1}(p_{n+1} - f^e_{n+1} - f^h_{n+1}) + ig\sqrt{n+1}\langle e^\dagger eh^\dagger h\ketbra{n+1}{n+1} - \underbrace{e^\dagger eh^\dagger h\ketbra{n}{n}}_{C_n^X}\rangle\label{eq:der_psin}.
     \end{split}
     \end{align}
The EoM for the semiconductor JCM can be closed by calculating the derivative of $C_n^X$, which reads
    \begin{align}
    \begin{split}
    \diff_t (e^\dagger e h^\dagger h \ketbra{n}{n})	 =& \mean{i[H, e^\dagger e h^\dagger h \ketbra{n}{n}]} \\
      &-ig \mean{h e e^\dagger e h^\dagger h b^\dagger\ketbra{n}{n} + \underbrace{\cancel{e^\dagger h^\dagger  e^\dagger e h^\dagger h b\ketbra{n}{n}}}_{\propto ~h^\dagger h^\dagger} - \underbrace{\cancel{e^\dagger e h^\dagger h hc \ketbra{n}{n}}b^\dagger}_{\propto ~h h} - e^\dagger e h^\dagger h  e^\dagger h^\dagger\ketbra{n}{n}b} \\
	=&ig \mean{e e^\dagger e h h^\dagger h b^\dagger\ketbra{n}{n} + e^\dagger e e^\dagger h^\dagger h h^\dagger \ketbra{n}{n}b} \\
		=&ig \sqrt{n+1}\mean{e h \ketbra{n+1}{n} + e^\dagger h^\dagger \ketbra{n}{n+1}} \\
		=&-ig \sqrt{n+1}\mean{h e\ketbra{n+1}{n} - \mathrm{h.c.}} \\
		=&2g\sqrt{n+1}~\psi_n,
    \end{split}
    \end{align}
and couples only to the already known polarization $\psi_n$.
\end{widetext}

\subsection{Parameter space and Correlation}
\label{app:parameter}
To capture the full range of possible initial electronic configurations of the QD in the semiconductor JCM we need not only the charge $C$, and the oscillation ability $O$, but also the QDs inversion $I$ in terms of the expansion coefficients of the density matrix $c_i$.
To this end we consider the following transformation:
\begin{equation}
  \begin{aligned}
	O &= c_2 + c_3 			  &c_2 &= \frac{1}{2}(O-I) \\
	C &= 1 - c_2 - c_3 - 2c_4 \quad  &c_3 &= \frac{1}{2}(O+I) \\
	I &= c_3 - c_2			  &c_4 &= \frac{1 - O - C}{2}.
  \end{aligned}
\end{equation}
The coordinates are bounded by the values
\begin{align}
	C \in [-1,1] \quad O \in [0,|C|] \quad I \in [-O,O], \label{eq:domain}
\end{align}
which is reflected by the triangular shape of the $g^{(2)}_{max}(0)=f(O,C)$ plot in Fig.~\ref{fig:correlation}(b). To specify the number of excitations in the QD, we have chosen the initial condition $I=0$, i.e.,~$c_2=c_3$.

\section{Dephasing}

\subsection{Analytical solution}
\label{app:deph}
Since we are interested in the long term effect of the non-local dephasing without s-shell decay, $\beta=0$, only six of the eight equations of Eq.~(\ref{eq:diffconfig}) have to be taken into account. The matrix $M$ simplifies to
\begin{align}
	M=
    \left(\begin{array}{c|c}
		W(\Gamma) & \mathbf{0} \\\hline
		G		  & W(0)
    \end{array}\right),
\end{align}
where $W$ and $G$ are defined by
\begin{align}
    W(\Gamma) &= \left(\begin{array}{ccc}
         	-\Gamma& \omega&    \sz   \\
        	-\nicefrac{\omega}{2} & -\Gamma& \nicefrac{\omega}{2}  \\
            \sz&  -\omega&   -\Gamma
    \end{array}\right), \quad
    G = \left(\begin{array}{ccc}
    		\Gamma&  \sz&    \sz \\
        	\sz&  \{\Gamma, 0\}&    \sz   \\
            \sz&  \sz&    \Gamma
    \end{array}\right)\nonumber,
\end{align}
with the notation $\omega = 2g$.

In the many-particle basis (formulation (ii)) , the solution of the system with the initial condition $r_0 = (1,0,0,0,0,0)^T$ reads
\begin{align}
   \left(\begin{array}{c}
        X\!X^0\\
        \psi_X^0\\
        X_p^1\\
        X_s^0\\
        \psi^0\\
        G^1\\
    \end{array}\right)=
	T\left(\begin{array}{cc}
		\nicefrac{1}{2} & 1\\
		-\frac{1}{8}\left(\frac{\Gamma}{\Gamma + 2i\omega}+1\right) & e^{i\omega t} \\
		-\frac{1}{8}\left(\frac{\Gamma}{\Gamma - 2i\omega}+1\right) & e^{-i\omega t} \\
		\nicefrac{1}{2} & e^{-\Gamma t}\\
		\nicefrac{1}{8} & e^{-\Gamma t} e^{-i\omega t} \\
		\nicefrac{1}{8} & e^{-\Gamma t} e^{-i\omega t}
    \end{array}\right),\label{eq:sol6}
\end{align}
where $T$ is the transformation matrix, consisting of all eigenvectors of $M$. For $\Gamma \neq 0$, the asymptotic behavior is determined by the first three rows of Eq.~(\ref{eq:sol6}). For $X_s^0(t+\tau)$, with $\tau \rightarrow \infty$, we obtain
\begin{align*}
	X_s^0(t)|_{t\gg 0} = \frac{(2\Gamma^2+4\omega^2)\cos{\omega t}+2\Gamma\omega\sin{\omega t}}{4(\Gamma^2+4\omega^2)}+ \frac{1}{2} \nonumber \\
	= \frac{\sqrt{(2\Gamma^2+4\omega^2)^2 + (2\Gamma\omega)^2}}{4(\Gamma^2+4\omega^2)}\sin{(\omega t + \varphi)} + \frac{1}{2},
\end{align*}
where $\varphi$ is an irrelevant phase. The oscillation of $X_s^0$ is centered around $\nicefrac{1}{2}$, and its amplitude varies from $\nicefrac{1}{4}$ for small, but nonzero dephasing rates $\Gamma$, to $\nicefrac{1}{2}$, for great $\Gamma \gg \omega$.

In the single-particle perspective (formulation (i)), the solution reads
\begin{align}
   \left(\begin{array}{c}
        X\!X^0\\
        \psi_X^0\\
        X_p^1\\
        X_s^0\\
        \psi^0\\
        G^1\\
    \end{array}\right)=
	T\left(\begin{array}{cc}
		\nicefrac{1}{2} & 1\\
		-\nicefrac{1}{4} & e^{i\omega t} \\
		-\nicefrac{1}{4} & e^{-i\omega t} \\
		\nicefrac{1}{2} & e^{-\Gamma t}\\
		\nicefrac{1}{4} & e^{-\Gamma t} e^{i\omega t} \\
		\nicefrac{1}{4} & e^{-\Gamma t} e^{i\omega t}
    \end{array}\right)
\end{align}
and the coefficients do not depend on $\omega$ or $\Gamma$. Therefore, the amplitude of
\begin{align}
	X_s^0|_{t\gg0} = \frac{1}{2}(\sin{(\omega t + \delta)} + 1)
\end{align}
stays $\nicefrac{1}{2}$ for all values of $\Gamma$ and $\omega$.

\subsection{Pumped p-exciton}
\label{app:pump}
In Sec.~\ref{ssec:ssd} in the main text we have presented a minimal example that induces the non-local dephasing effect. To demonstrate that the dephasing can become significantly stronger we present a further exploitation of the non-local dephasing mechanism.

\begin{figure}[h]
	\centering
    \includegraphics[width=0.48\textwidth]{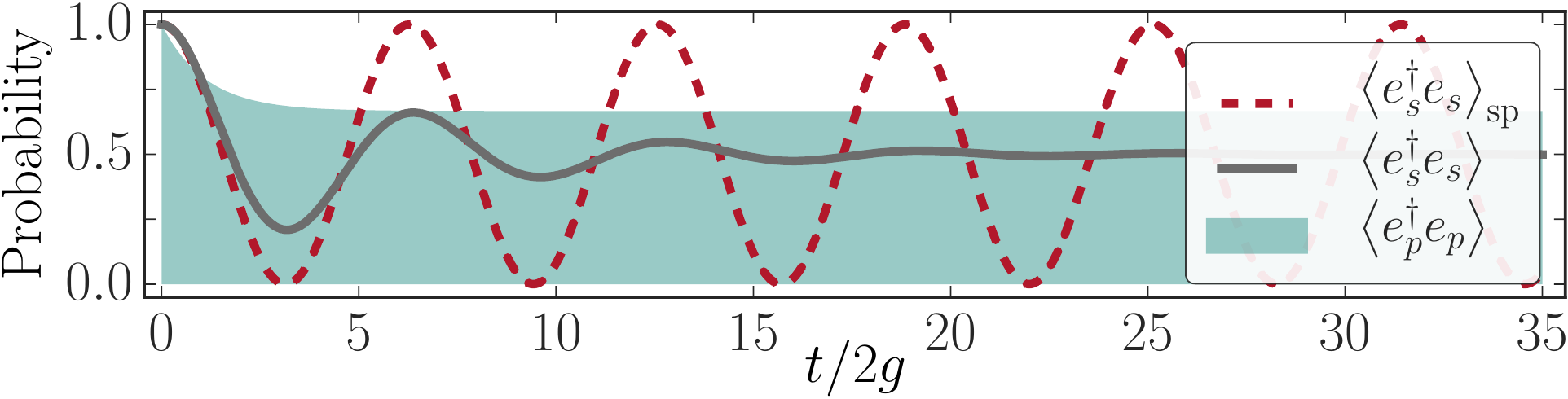}
  \caption{Dynamics of $\mean{e^\dagger_se_s}$ in the single-particle basis (red dashed line) and in the many-particle basis (black line), and of $\mean{e^\dagger_pe_p}$ (shaded area), for the same parameters as in Fig.~\ref{fig:dephasing}(c) with additional p-shell pump $P=0.3=\Gamma$.}
  \label{fig:mitpumpe}
\end{figure}

When we add a pumping process to the p-exciton with rate $P$ to our model in Sec.~\ref{ssec:ssd} we obtain a case where different constructions of the dissipator (i) and (ii) result in an entirely different long term behaviour of the system.
The pumping process is induced by the adjunct collapse operators for the p-exciton decay. In the single-particle basis (i) this is
$L_{Psp}=e^\dagger_p h^\dagger_p=\ketbra{X_p}{G} + \ketbra{X\!X}{X_s}$  (adjoint of Eq.~(\ref{eq:Lsp}), and in the configuration basis (ii) these operators are $L_{PG}=\ketbra{X_p}{G}$ and $L_{PX}=\ketbra{X\!X}{X_s}$  (adjoint of Eq.~(\ref{eq:LGLX})). As shown in Fig.~\ref{fig:mitpumpe}, the s-shell electron occupation $\langle e^\dagger_se_s \rangle$ reaches a steady state for a dissipator constructed in the configuration basis, whereas in the single-particle basis the occupation $\langle e^\dagger_se_s\rangle_{\text{sp}}$ performs Rabi-oscillations for all times, independent of the p-shell decay and pumping.

\end{document}